\documentclass[showpacs,superscriptaddress,twocolumn,floatfix]{revtex4}

\usepackage{amsmath}
\usepackage{amssymb}
\usepackage{bm}
\usepackage{graphicx}

\begin{document}

\title{
Manipulating optical signals at sub-wavelength scale by planar
arrays of metallic nanospheres: Towards plasmonic interference
devices}

\author{A.\ V.\ Malyshev}
\thanks{On leave from Ioffe Physiko-Technical Institute, 26
Politechnicheskaya str., 194021 St. Petersburg, Russia }
\affiliation{ Centre for Theoretical Physics and Zernike Institute
for Advanced Materials, University of Groningen, Nijenborgh 4, 9747
AG Groningen, The Netherlands }
\affiliation{ GISC, Departamento de F\'{\i}sica de Materiales,
Universidad Complutense, E-28040 Madrid, Spain }

\author{V.\ A.\ Malyshev}
\affiliation{ Centre for Theoretical Physics and Zernike Institute
for Advanced Materials, University of Groningen, Nijenborgh 4, 9747
AG Groningen, The Netherlands }

\author{J.\ Knoester}
\affiliation{ Centre for Theoretical Physics and Zernike Institute
for Advanced Materials, University of Groningen, Nijenborgh 4, 9747
AG Groningen, The Netherlands }

\begin{abstract}
We show that interference can be the principle of operation of an
all-optical switch and other nanoscale plasmonic interference
devices (PIDs). The optical response of two types of planar
plasmonic waveguides is studied theoretically: bent chains and
Y-shaped configurations of closely-spaced metallic nanospheres. We
study symmetric Y-shape arrays as an example of an all-optical
switch and demonstrate that effective phase- and amplitude-sensitive
control of the output signal can be achieved due to interference
effects.
\end{abstract}


\pacs{
    78.67.-n    
    73.20.Mf    
    85.35.-p    
}

\maketitle

Sub-wavelength guiding of optical signals is of great interest to
information transport and processing. New horizons were opened when
linear arrays of metal nanoparticles were introduced as wave guides
\cite{Quinten98}. It was shown that optical excitations can be
channeled into these nanometer scale systems and transmitted over a
distance on the order of 1000 nm. Such waveguides are nowadays known
as plasmonic arrays. The early work was followed by many other
studies of optical guiding at the sub-wavelength
scale~\cite{Brongersma00,Hernandez05,Waele07,Markel07,Malyshev08,Maier07}.
The redirection of optical signals at the nanometer scale is an even
greater challenge. To this end, researchers have studied the
channeling of surface plasmon polaritons by sub-wavelength V-shaped
grooves on a metal surface~\cite{Pile05,Bozhevolnyi05}. Y-splitters,
Mach-Zender interferometers, and waveguide-ring resonators have been
realized by using bent grooves~\cite{Bozhevolnyi06}, as well as
dielectric-loaded surface plasmon polariton
waveguides~\cite{Holmgaard08,Bozhevolnyi08}. Also plasmonic arrays
have been considered as Y or T shaped splitters, with the switching
induced by the interaction with a nearby molecule \cite{Neuhauser07}
or the phase and polarization of the light in combination with
genetic algorithms \cite{Sukharev06}.

In this Letter, we focus on manipulating optical signals at the
nanometer scale using two types of planar plasmonic arrays of
metallic nanoparticles: bent chains and Y-shaped configurations of
identical nanospheres. We show that, in contrast to sharply bent
chains, those with a smooth bend can redirect optical signals over
an arbitrary angle, up to $90^\circ$. The Y-shaped geometry presents
two options for controlling the flow of electromagnetic energy: (i)
splitting the incoming flux into two or (ii) merging two signals
from different branches into one. The efficiency of splitting turns
out to be not very high. In the case of the mixer, the relative
phase of one (control) signal with respect to the other (input) one
allows for interference-induced phase and amplitude control of the
output. This opens perspectives for designing nanoscale all-optical
switches based on interference phenomena.

The arrays we consider consist of several tens of identical silver
nanospheres of radius $a$ embedded in a homogeneous dielectric host
with a dielectric constant $\varepsilon_b = 2.25$, typical for
experiments with silver nanoparticles (see, e.g.,
Ref.~\onlinecite{Waele07}). The center-to-center distance $d$
between adjacent nanoparticles is chosen such that $a/d \le 1/3$, in
which case the point dipole approximation for the inter-particle
interactions can be used~\cite{Kelly03,Haynes03,Yong04}.

To calculate the response of the array to optical fields, we use the
quasi-static approximation and characterize the nanospheres by their
frequency-dependent polarizability $\alpha(\omega)$, taken in the
form
\begin{equation}
    \label{alpha}
    \frac{1}{\alpha} = \frac{1}{\alpha^{(0)}}
    -\frac{k^2}{a}
    -\frac{2}{3} i k^3 .
\end{equation}
Here, $\alpha^{(0)}$ is the bare polarizability of a single sphere
and the next two ($k$-dependent) terms account for the
depolarization shift and radiative damping,
respectively~\cite{Meier83}; $a$ stands for the sphere's radius,
while $k=\sqrt{\epsilon_b}\;\omega/c$ is the wavevector of the
exciting light in the host medium. The bare polarizability
$\alpha^{(0)}$ is expressed as
\begin{equation}
    \label{alphaLL}
    \alpha^{(0)} = a^3\; \frac{\varepsilon - \varepsilon_b}
    {\varepsilon + 2\,\varepsilon_b} \ ,
\end{equation}
where the dielectric constant of the metal is taken in the generalized
Drude form~\cite{Waele07}
\begin{equation}
    \label{varepsilon}
    \varepsilon = \epsilon - \eta\, \frac{\omega_p^2}{\omega^2
    + i\,\omega\,\gamma} \ .
\end{equation}
Here, $\omega_p$ and $\gamma$ are, respectively, the bulk metal's
plasma frequency and Ohmic damping constant, while $\epsilon$ and
$\eta$ are adjustable parameters. For silver, $\epsilon = 5.45$,
$\eta = 0.73$, $\omega_p = 1.72 \times 10^{16}$ rad/s, and $ \gamma
= 8.35 \times 10^{13}$ 1/s \cite{Waele07}.

We assume that the sphere at one end of the array (or the two end
spheres in the case of the mixer) is excited with a c.w. field of
frequency $\omega$, corresponding to a wave length $\lambda = 2\pi/k
= 2\pi c/(\omega\sqrt{\epsilon_b})$. We chose the polarization of
the exciting field perpendicular to the plane of the array, so that
only transverse plasmon modes will be excited in the array. Under
steady state conditions, the set of coupled equations for the
induced dipole moments ${p}_n$ of the spheres (labeled by $n$)
reads:
\begin{equation}
    \label{eqs}
    \sum_{m}\left( \frac{1}{\varepsilon_b\,\alpha}\,\delta_{nm}
    - G_{nm}\right)\,p_m = E_n\ .
\end{equation}
Here, $E_n$ is the incident field at the position ${\bf r}_n$ of the
$n$th dipole and $G_{nm}$ is the retarded Green's function of the
electric field in the dielectric background. For the geometry at
hand the latter is given by \cite{Weber04}
\begin{equation}
    \label{G}
    G_{nm}  =  \frac{1}{\varepsilon_b} \left(
    -\frac{1} {r_{nm}^3} + \frac{i k} {r_{nm}^2}
    + \frac{k^2} {r_{nm}} \right) e^{ikr_{nm}} \ ,
\end{equation}
where $r_{nm} = |{\bf r}_n - {\bf r}_m|$ and $G_{nn} = 0$.
Throughout the paper we consider planar arrays of silver nanospheres
with radii of 50 nm and center-to-center distances of 150 nm.

\begin{figure}[b]
\begin{center}
    \includegraphics[width=0.48\columnwidth]{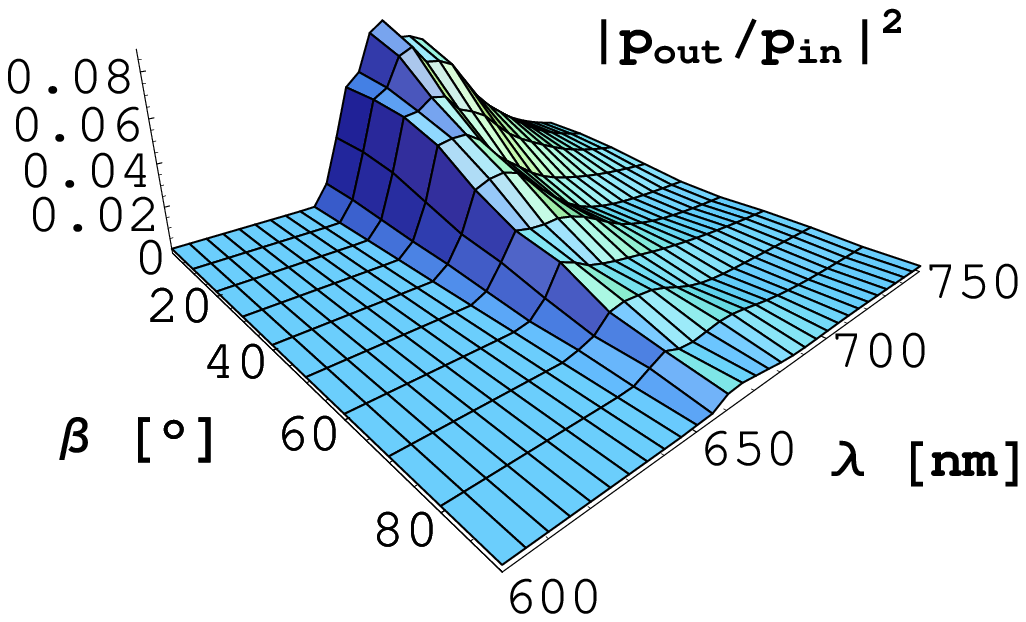}
    \includegraphics[width=0.48\columnwidth]{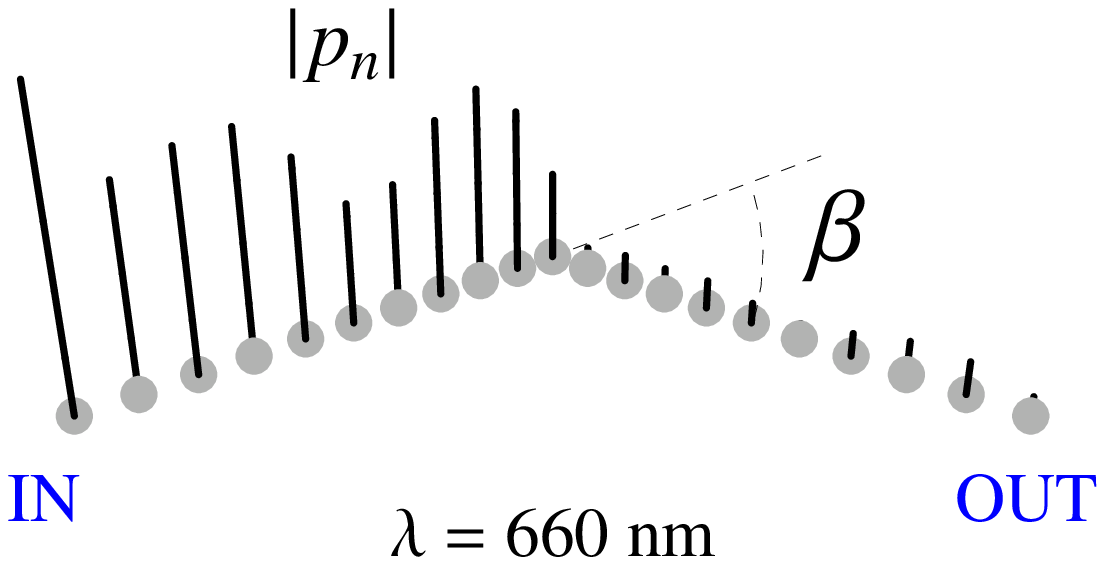}\\
    \includegraphics[width=0.48\columnwidth]{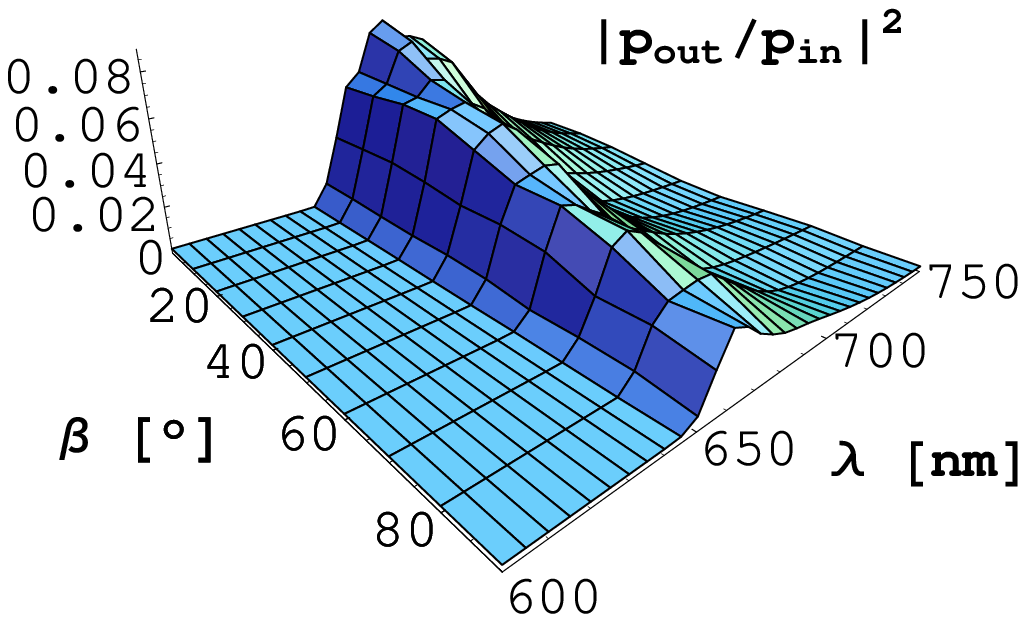}
    \includegraphics[width=0.48\columnwidth]{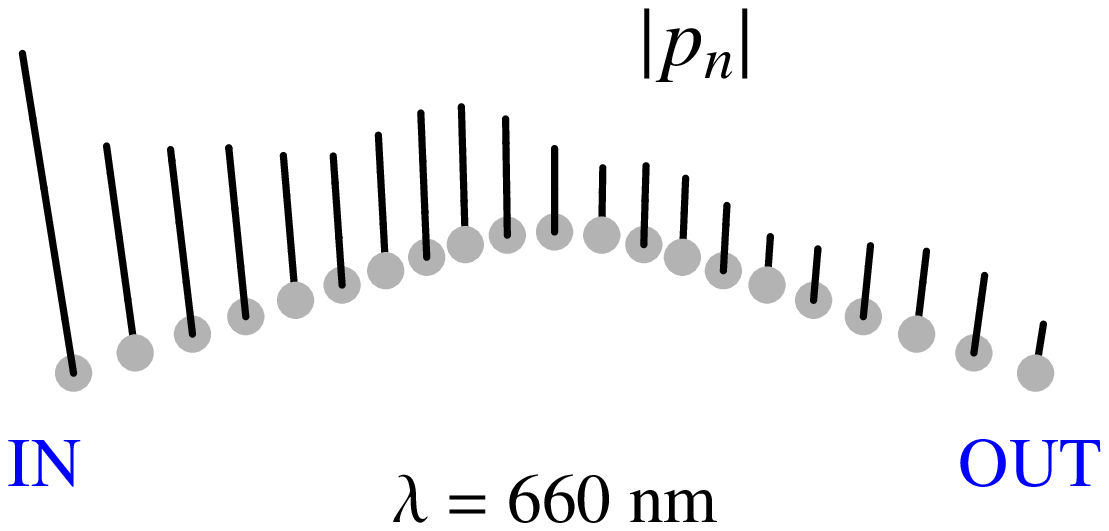}
\end{center}
    \caption{
    Optical signal in a chain of $N=21$ silver nanospheres
    with a sharp (top) and a smooth (bottom) bend.
    To the left is shown the dependence of the transmittance
    $T = |p_\mathrm{out}/p_\mathrm{in}|^2$
    on the wavelength $(\lambda)$ and the bending angle $(\beta)$;
    to the right schemes of the chains are displayed, together with the
    distributions of the magnitudes of the dipole moments $|p_n|$
    (indicated by the vertical sticks)
    for the case of $\beta=90^\circ$ and $\lambda= 660$ nm. }
\label{fig1}
\end{figure}

Figure~\ref{fig1} presents our results on redirecting optical
signals by bent plasmonic arrays. The top plots correspond to sharp
bends, where the chain abruptly changes direction over an angle
$\beta$, while the bottom plots present the case of smooth bends
consisting of three spheres (i.e. two spheres lie off the straight
lines formed by the input and output legs). In both cases, the chain
consists of $N = 21$ nanospheres and the leftmost sphere is driven
by the incoming field: $E_n=E_0\delta_{n1}$. To the left is shown
the transmittance $T = |p_\mathrm{out}/ p_\mathrm{in}|^2 = |p_{N}/
p_{1}|^2$ as a function of $\beta$ and $\lambda$. Clearly, for both
cases the maximum transmission occurs around $\lambda = 660$ nm;
however, for bending angles larger than $45^\circ$, the transmission
through the smooth bend exceeds by far the one for the sharp bend.
This is also seen from the plots to the right, which for both cases
show the amplitudes $|p_n|$ of the dipoles for a bending angle
$\beta = 90^\circ$. The explanation is that a smooth bend reduces
reflection and radiation losses.

\begin{figure}[t]
\begin{center}
    \includegraphics[width=0.48\columnwidth]{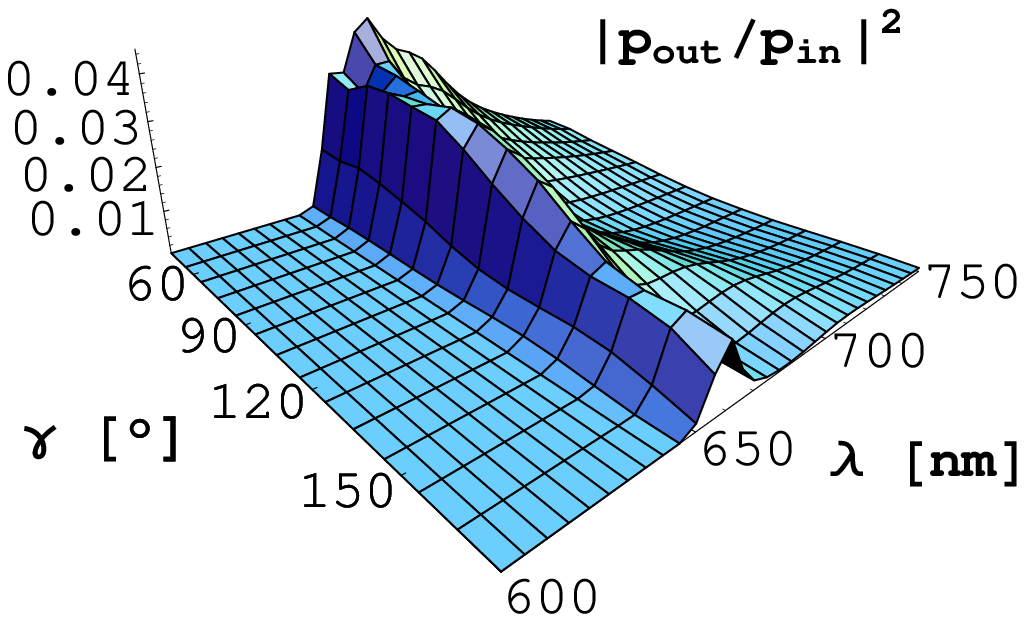}
    \includegraphics[width=0.48\columnwidth]{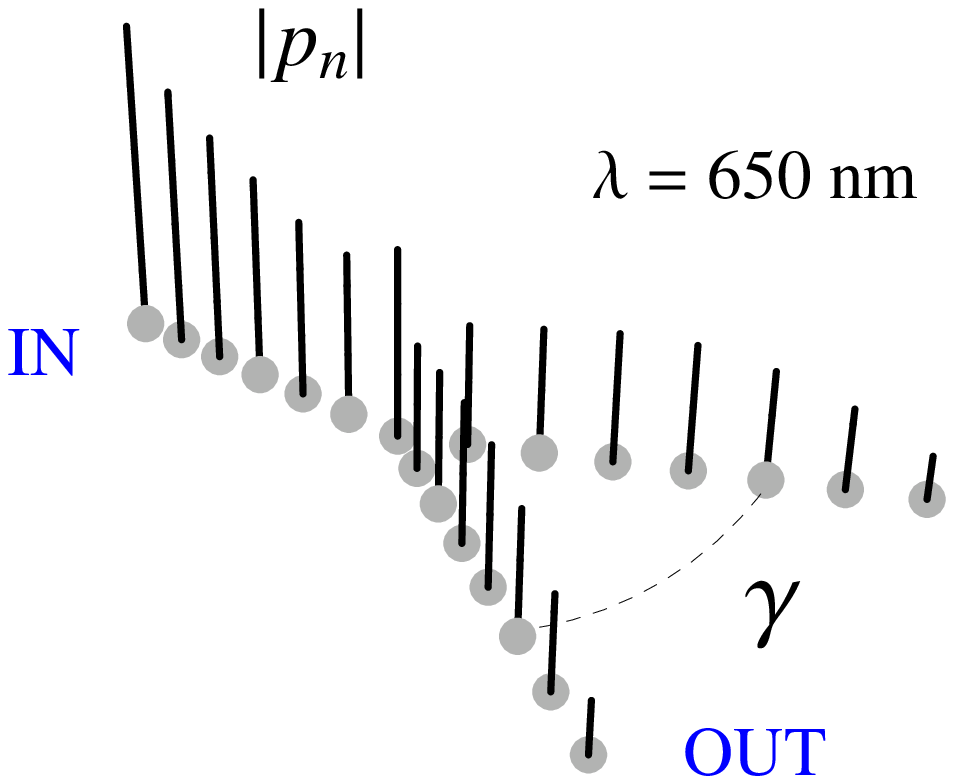}
\end{center}
\caption{
    Y-shaped splitter with seven nanospheres in each
    branch. The left-most sphere is driven by the incoming field.
    To the left is shown the dependence of the transmittance
    on the wavelength ($\lambda$) and the aperture angle ($\gamma$). To the right a scheme of the
splitter is displayed,
    together with the distribution of the dipole moments $|p_n|$
    (vertical sticks)
    for the case of $\gamma=60^\circ$ and $\lambda= 650$ nm.
}
\label{fig2}
\end{figure}

We now turn to Y-shaped arrays. First, we discuss the efficiency of
splitting one signal into two. Figure~\ref{fig2} displays our
results for a symmetric Y-shaped splitter comprised of seven
nanospheres in each branch and with variable aperture angle
$\gamma$. The left-most sphere (denoted IN in Fig.~\ref{fig2}) is
driven by the incoming field, while the output is measured at the
end sphere of one of the two other branches (denoted OUT). To the
left is displayed the dependence of the transmittance
$T=|p_\mathrm{out}/p_\mathrm{in}|^2$ on $\gamma$ and $\lambda$. To
the right the Y-splitter is shown with the spatial distribution of
the $|p_n|$, calculated for $\gamma=60^\circ$ and the wavelength
$\lambda = 650$ nm, at which the output $|p_\mathrm{out}|$ is
maximum. We see that a reasonable fraction of the signal is
transmitted through the branching point; nevertheless, the overall
transmission to the output point is low, mostly due to radiative
losses.

\begin{figure}[t]
\begin{center}
\includegraphics[width=0.48\columnwidth]{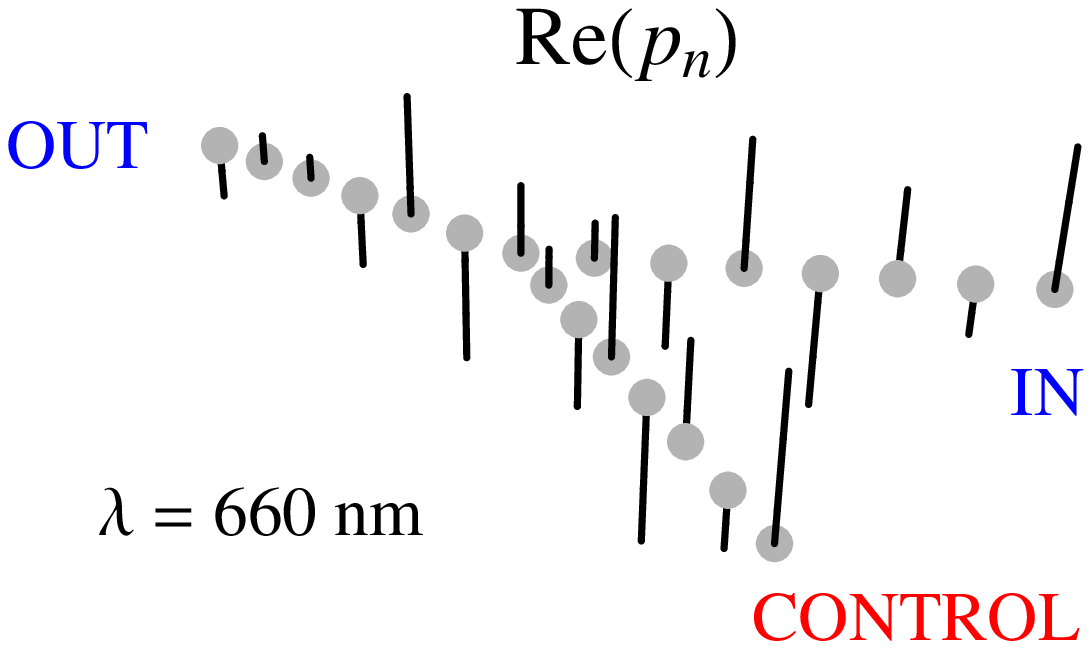}
\includegraphics[width=0.48\columnwidth]{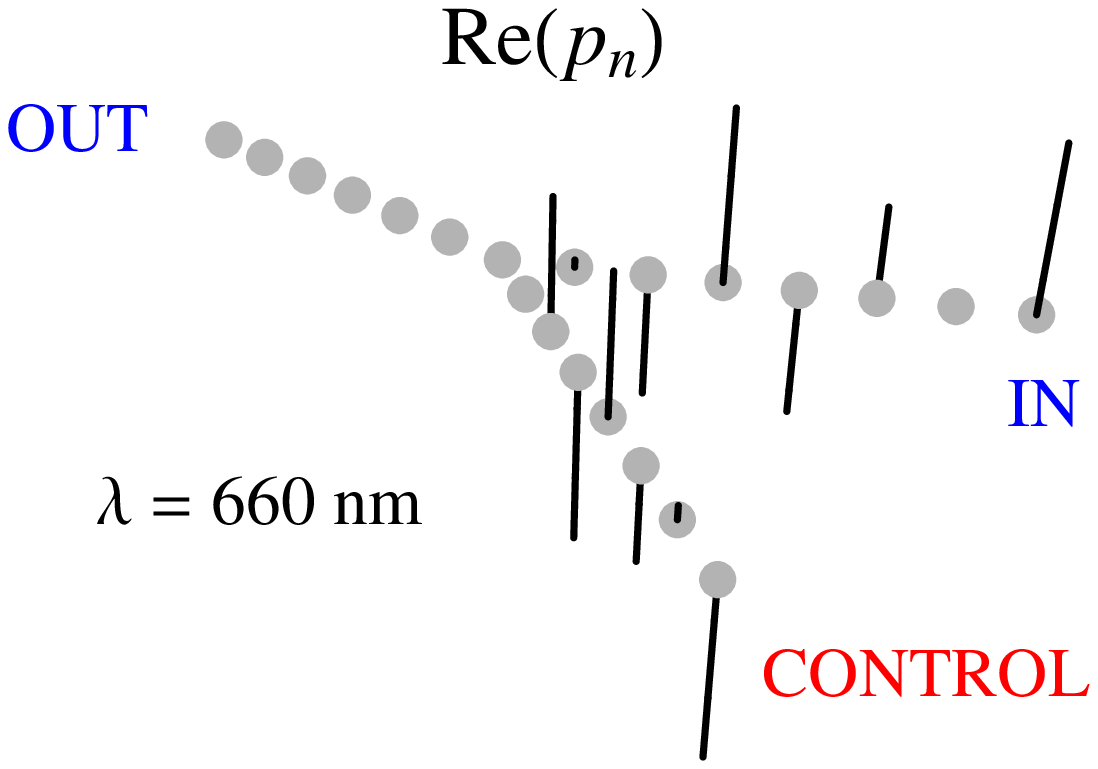}
\\
\includegraphics[width=0.48\columnwidth]{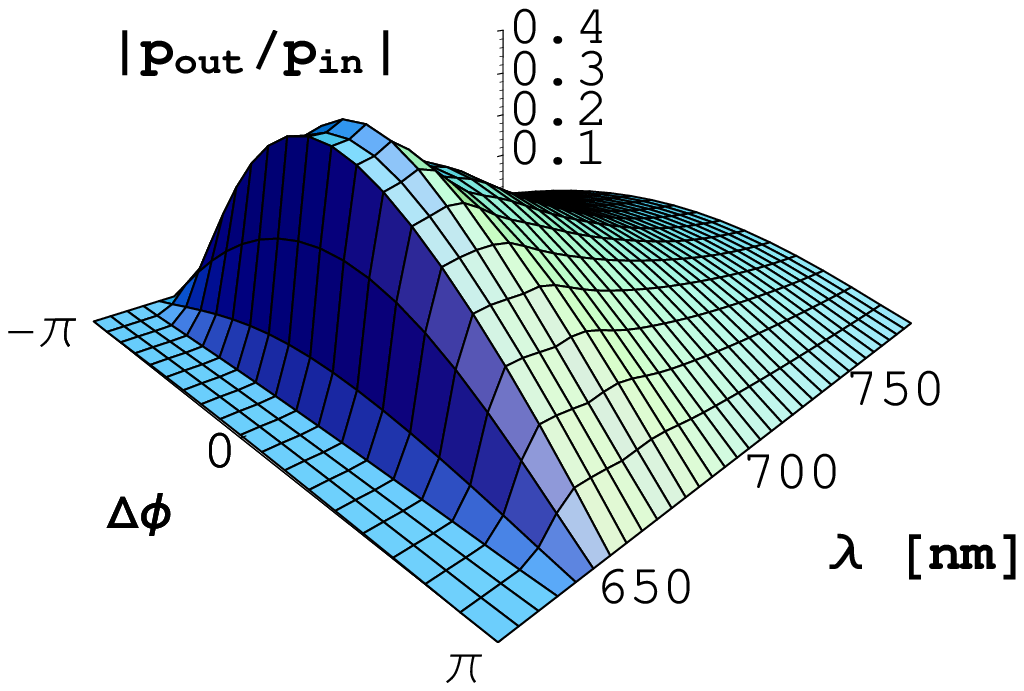}
\includegraphics[width=0.48\columnwidth]{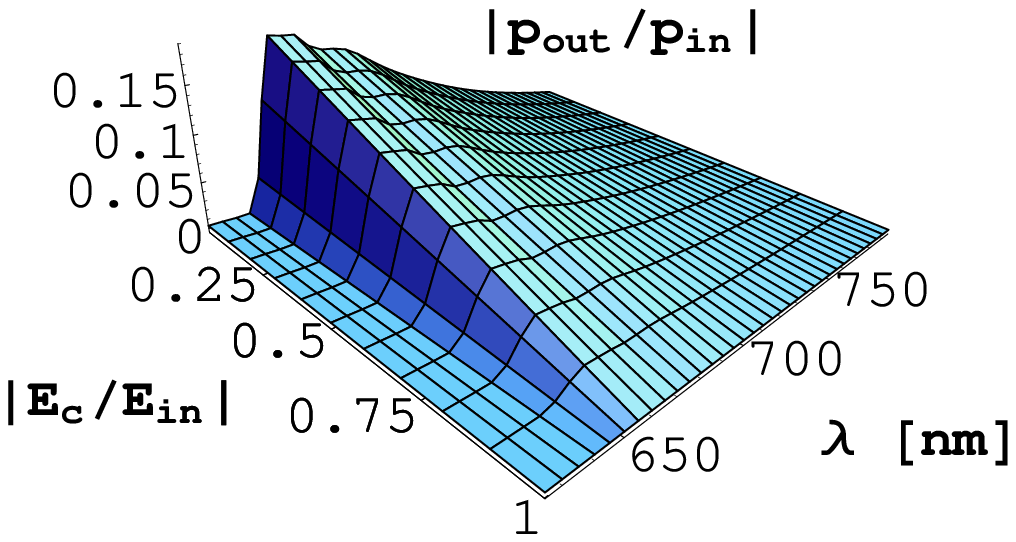}
\end{center}
    \caption{
    Y-shaped mixer with seven nanospheres in each branch and an aperture
    angle $\gamma = 60^\circ$.  The two right-most spheres (CONTROL and IN)
    are driven by the incoming fields. Upper plots --- spatial distributions
    of $Re(p_n)$ (vertical sticks), calculated for a phase difference between
    the driving fields of $\Delta\phi = 0$ (left) and $\Delta\phi = \pi$
    (right). Lower left-hand plot --- $(\Delta\phi,\lambda)$-dependence of
    the amplitude transmittance $t = |p_\mathrm{out}/p_\mathrm{in}|$ for the
    case of equal driving field amplitudes. Lower right-hand plot ---
    dependence of the amplitude transmittance on $\lambda$ and on the ratio
    $|E_c/E_{in}|$ of the amplitudes of the control an input fields for
    $\Delta\phi = \pi$.
    %
    %
    }
\label{fig3}
\end{figure}

Figure~\ref{fig3} demonstrates all-optical signal control using the
Y-shaped plasmonic array described above. Here we excite the extreme
spheres of two branches of the array, called IN and CONTROL in the
upper panels of Fig.~\ref{fig3} and focus on the output signal
$p_{\mathrm{out}}$ at the other end of the mixer. In the top panels
we present the spatial distribution of $Re(p_n)$ calculated for the
case when the input and control signals have the same amplitude and
either are in phase (left) or out of phase (right). The calculation
was performed for the wavelength $\lambda = 660$ at which the output
$|p_{\mathrm{out}}|$ is maximal for the in-phase case. If the phase
difference $\Delta\phi$ between the input and control signals
vanishes, they interfere constructively, resulting in a non-zero
output. In the opposite case, $\Delta\phi = \pi$, destructive
interference leads to zero output. Clearly the output can be
controlled by changing $\Delta\phi$. The lower left-hand panel in
Fig.~\ref{fig3} presents the overall dependence of the amplitude
transmittance $t = |p_\mathrm{out}/p_\mathrm{in}|$ on $\Delta\phi$
and $\lambda$. For the symmetric mixer considered, $t\propto
|\cos(\Delta\phi/2)|$ as we will show below.

Another way of all-optical control is achieved by keeping the phase
difference fixed at $\Delta\phi = \pi$, while varying the amplitude
of the control field. For this situation, the lower right-hand panel
in Fig.~\ref{fig3} shows the amplitude transmittance $t$ as a
function of $\lambda$ and the relative control amplitude
$|E_\mathrm{c}/E_\mathrm{in}|$. The dependence is linear in
$|E_\mathrm{c}/E_\mathrm{in}|$, reminiscent of the traditional
electronic transistor. When the control amplitude equals the input
one, the output signal turns to zero.

Both modes of optical control demonstrated above, phase and
amplitude control, follow from the linear nature of the response
(superposition), combined with the system's symmetry. As a result,
the signal in the output branch of a symmetric Y-shaped mixer is
proportional to $E_\mathrm{in}+E_\mathrm{c}$. In the case of equal
input and control amplitudes,
$E_\mathrm{c}=E_\mathrm{in}e^{i\Delta\phi}$, yielding
$|p_{\mathrm{out}}|\propto|\cos(\Delta\phi/2)|$. On the other hand,
for $\Delta \phi = \pi$, one gets
$|p_{\mathrm{out}}|\propto|1-E_\mathrm{c}/E_\mathrm{in}|$. These
dependencies on $\Delta\phi$ and $E_\mathrm{c}/E_\mathrm{in}$ are
indeed observed in Fig.~\ref{fig3}. Thus, Y-shaped plasmonic
waveguides provide for the possibility of either phase- or
amplitude-sensitive all-optical control of signals at the
sub-wavelength scale. We stress that for a complete cancelation of
the input signal by the control one over a wide range of
wavelengths, the system should be symmetric~\cite{note1}.

In conclusion, we studied two types of planar plasmonic waveguides:
bent chains and Y-shaped configurations of silver nanospheres. We
found that: (i) a smooth bend can redirect optical signals much
better than a sharp bend with the same turning angle (ii) Y-shaped
arrangements of nanospheres can operate as a splitter of optical
signals or as an effective mixer. In the latter case, the output may
be controlled by tuning the phase of the control signal relative to
that of the input signal. In this regime the system operates as a
phase-dependent all-optical switch. Alternatively, the control
signal can be kept out of phase relative to the input signal, while
tuning its amplitude. In this case the device operates as an optical
transistor with a linear control-output characteristic. Finally, we
stress that the principles underlying the mixer presented here are
very general and also apply to other plasmonic (or radio-frequency)
systems. We therefore believe that plasmonic interference devices
(PIDs) similar to those considered here, can be fabricated using a
wide range of available techniques, such as V-shaped grooves on
metal surfaces~\cite{Bozhevolnyi06}, dielectric-loaded surface
plasmon polariton waveguides~\cite{Holmgaard08}, etc.

This work was supported by NanoNed, a national nanotechnology programme
coordinated by the Dutch Ministry of Economic Affairs.


\begin{thebibliography}{99}
\bibitem{note1} The possibility of such control in the microwave range by
T-shaped structures was also discussed in Ref.~\cite{Maier01}, but
no complete cancelation of the output was achieved, due to the lack
of symmetry.
\end{thebibliography}
\end{document}